\documentclass[a4paper]{VisionStyle}
\usepackage{epsfig}

\begin{document}

\title{Luminosity functions of X-ray binaries
       and their dynamical history of the host galaxies }

\author{K.\,Wu\inst{1,}\inst{3} \and
        A.F.\,Tennant\inst{2} \and
        D.A.\,Swartz\inst{2} \and
        K.K.\,Ghosh\inst{2} \and
        R.W.\,Hunstead\inst{3} }

\institute{
  Mullard Space Science Laboratory, University College London,
  Holmbury St Mary, Surrey, RH5 6NT, United Kingdom
\and
  SD-50, NASA Marshall Space Flight Center, Huntsville,
  AL 35812, USA
\and
  School of Physics A28, University of Sydney, NSW 2006, Australia}

\maketitle

\begin{abstract}

The X-ray sources observed in nearby galaxies
  show brightness distributions (log($N(>S)$)--log($S$) curves)
  which can be described as single or broken power laws.
Single power-law distributions are often found for sources in galaxies
  with vigorous ongoing star-formation activity,
  while broken power laws are found in elliptical galaxies
  and in bulges of spiral galaxies.
The luminosity break can be caused by a population of X-ray binaries
  which contain a neutron star accreting at the Eddington limit
  or by aging of the X-ray binary population.
We show that a simple birth-death model
  can reproduce single and broken power-law
  log($N(>S)$)--log($S$) curves.
We have found that power-law log($N(>S)$)--log($S$) curves
  are a consequence of smooth continuous formation of X-ray binaries
  and a luminosity break is the signature of a starburst episode
  in the recent past.
The luminosity break is robust
  and is determined by the lifespans of the X-ray binaries
  relative to the look-back time to the starburst epoch.
The model successfully explains the different forms
  of log($N(>S)$)--log($S$) curves for the disk and bulge sources
  in the spiral galaxy M81.

\end{abstract}

\section{Introduction}

Recent Chandra observations have shown that
  nearby spiral galaxies similar to the Milky Way 
  (e.g., M81, \cite{kwu-E1:ten01}) 
  contain hundreds of discrete X-ray sources.
The source numbers are even large 
  in giant ellipticals, starburst galaxies and mergers
  (see \cite{kwu-E1:sar00}, \cite{kwu-E1:mat01}, \cite{kwu-E1:fab01}).
While these sources are an inhomogeneous population,
  the majority of them are believed to be X-ray binaries (XRBs)
  containing a black hole or neutron star.
The X-rays from these binaries are powered by accretion,
  with the duration of their X-ray phases limited by
  the supply of accreting material from their companion stars,
  which are mostly main-sequence or giant stars.
The lifespans of XRBs are therefore finite.
If the averaged mass-transfer rates are higher for brighter systems,
  then brighter XRBs are short-lived.

For the X-ray luminosities of the Chandra sources in galaxies 
  at distances of a few Mpc
  ($L_{\rm x} > 10^{36}~{\rm erg~s}^{-1}$),
  the duration of their mass-transfer (X-ray) phase are shorter than $10^9$~yr. 
These timescales are short
  in comparison with the dynamical timescales of galaxies,
  which are usually longer than $10^9$~yr.
XRBs are thus formed and then cease to be X-ray active
  within the lifetime of their host galaxies.
As the formation of XRBs is a ongoing process
  like the formation of normal stars,
  XRBs populate not just starburst galaxies
  but galaxies of different ages and of different Hubble types.

If a galaxy is undisturbed so that its stellar population ages smoothly,
  then the formation of XRBs is also likely to be a smooth process.
If the galaxy has an episode of vigorous star-formation
  triggered by galactic interaction or merging,
  then enhancement in the rates of XRB formation will subsequently occur.
The XRB populations are therefore tracers
  of the dynamical history of galaxies.

Here, we construct a birth-death model
  and use it to describe the time-dependent evolution 
  of the XRB populations in galaxies.
We show that the model is able to qualitatively explain
  the morphology of X-ray luminosity functions of XRBs in nearby galaxies.
In particular, the model provides a mechanism for the formation
  of a luminosity break in the log($N(>S)$)--log($S$) curves of the XRBs
  (where $S$ is the observed X-ray flux).
The implications of the results from this study
  on our understanding of galactic dynamics and evolution
  are briefly discussed.

\section{Birth-death model}

For an XRB population in a galaxy
  with the number of systems
  within a luminosity range $(L,L+\Delta L)$ at time $t$
  specified by $n(L,t)\Delta t$,
  the cumulative luminosity distribution of XRBs is
\begin{equation}
  N(>L,t) = \int_L^\infty dL' ~n(L',t) \ .
\end{equation}
As the sources are more or less at the same distance $d$,
  we can adopt $N(>S,t) = N(>L,t)$,
  where the observed flux $S \propto L/4 \pi d^2$.
For a population of systems determined
  only by birth and death processes 
  (the beginning and end of the X-ray active phase, respectively),
  the evolution of $n(L,t)$ is governed by a linear differential equation
  of the form
\begin{equation}
  {d \over {dt}}\ n(L,t) = f(L,t) - g(L,t) \ ,
\end{equation}
   where $f(L,t)$ is the birth rate and $g(L,t)$ is the death rate.

\subsection{an idealised case}

As a simple illustration, consider the case in which
  (i) the sources have a finite X-ray phase duration
      $t_{\rm x}(L)$, say, proportional to $1/L$,
  (ii) the source brightnesses are constant throughout the X-ray phase,
  and (iii) the sources are born at the same time, $t_{\rm a}$.
We further consider a birth-rate function $f(L,t) =  
\lambda(L)\delta(t-t_{\rm a})$,
  where $\lambda(L)$ is the brightness distribution of the sources at birth, 
  and $\delta(..)$ is the usual Dirac $\delta$-function.
In an absolute deterministic case,
  the death-rate is a time translation of the birth rate:
  $g(L,t) = f(L,t-t_{\rm x}(L)) = \lambda(L)\delta(t-t_{\rm a}-t_{\rm x}(L))$.   

\begin{figure}[ht]
  \begin{center}
   \epsfig{file=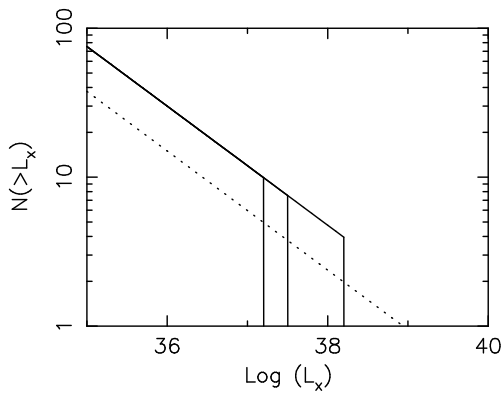, width=7.5cm}
  \end{center}
\caption{ 
   The starburst component of the $N(>L)$ curves 
     calculated from the idealised birth-death model (solid lines), 
     without differentiating black-hole and neutron-star binaries.   
   We have assumed that $t_{\rm x}(L) = (\beta L)^{-1}$, 
     where $\beta = [~0.3 M_\odot c^2~]^{-1}$,   
     and that that initial brightness distribution 
     $\lambda(L) \propto L^{-1.4}$,  
     such that the $N(>L)$ curves 
     have slopes similar to that of the Chandra sources in M81  
     (see Tennant et al.\ 2001).  
   Three cases, with $(t-t_{\rm a}) =$ 0.1, 0.5 and 1~Gyr, are shown.  
   Their luminosity cutoffs 
     are at $1.7 \times 10^{38}$, $3.4 \times 10^{37}$ and 
     $1.7 \times 10^{37}$~erg~s$^{-1}$ respectively.   
   The $N(>L)$ curve for the continuous star-formation component 
     is also shown for comparison. 
   We consider only the case 
     in which the slope of ${\bar n}_0(L)$ is the same as the slope of $\lambda(L)$ 
     and the normalisation is 1/2 of the normalisation of the starburst component.  
   }
\label{fauthor-E1_fig:fig1}
\end{figure}

When $f(L,t)$ and $g(L,t)$ are so specified, we can integrate equation (2)
  with respect to $t$, yielding
\begin{equation}
  n(L,t) =  {\bar n}_0(L) +
  \lambda(L)\theta(t-t_{\rm a})[1-\theta(t-t_{\rm a}-t_{\rm x}(L))]    \ ,   
\end{equation}
  where $\theta(..)$ are the Heaviside unit step-functions.
The constant ${\bar n}_0(L)$ is zero,
  because of the assumption that no systems are born except at $t_{\rm a}$.  
The result here is clear --- we only see XRB after a time $t_{\rm a}$,
  and we will not see systems with $L$ after
  a time $t_{\rm a}+t_{\rm x}(L)$.
Moreover, the bright systems disappear first, followed by fainter systems.   
If $t_{\rm x}(L)=(\beta L)^{-1}$,
  where $\beta$ is a constant,
  then a step-like luminosity cutoff is formed
  at $L_{\rm c} = [\beta (t-t_{\rm a})]^{-1}$ in $n(L,t)$
  and in the cumulative luminosity distribution $N(>L,t)$.
As the XRB population ages, 
  the cutoff migrates from high to low luminosities (Fig.~1).

Galaxies also have continuous formation of XRBs
  due to ongoing star-formation
  (such as young high-mass XRBs in the arms of spiral galaxies)
  or orbital and nuclear evolution of binary systems
  (such as old low-mass XRBs 
  in elliptical galaxies or in bulges of spiral galaxies).
This will produce a population of XRBs
  in addition to those formed as a result of starbursts.
If $t$ is large such that the birth and death of the XRBs due to continuous
  formation
  has reached an equilibrium, such that $dn(L,t)/dt|_{\rm con} = 0$,
  the linearity of the formulation allows us to simply set a non-zero ${\bar n}_0(L)$ 
  in equation (3) and obtain a solution immediately.
Provided that the XRB population due to the starburst dominates,
  the luminosity cutoff should be evident (Fig.~1).

The formulation can be easily generalised
  to take into account the delay in the XRB formation
  after a starburst episode at $t_{\rm a}$.
We can parametrise the delay $\epsilon(L)$ as a function $L$,  
  such that $f(L,t) = \lambda(L) \delta(t - (t_{\rm a} +\epsilon(L)))$
  and $g(L,t) = \lambda(L) \delta(t - (t_{\rm a} +\epsilon(L))-t_{\rm x}(L))$.  
It is again straightforward to integrate equations (2) and (1) directly
  and show qualitatively that the shape of the $N(>L)$ distribution of the XRBs 
  obtained in the idealised case still hold.
The only difference is that
  now the luminosity cutoff is no longer step-like.
A more realistic and comprehensive modeling of the delay effects
  will be presented in Wu et al.\ (in preparation).

\subsection{a stochastic model}

The purely deterministic approach given above needs to be modified
  because the onset of the X-ray phases of XRBs cannot be simultaneous
  and the systems may undergo transitions between bright and faint states. 
The observed luminosities $L$ are not simply 
  the lifespan-averaged luminosities
  which can be related to the duration of the X-ray active phase.
Also, the lifespans of the sources are characterised by their internal clocks, 
  which are the individual source ages $\tau$,
  and they are not the same as the external clock
  by which the time parameter $t$ specifying the evolution
  of the XRB population has been defined.
The ages of individual systems are not direct observables
  in the study of populations of XRBs in galaxies.
Only globally-averaged properties such as luminosity distributions
  can be measured.
Hence, the death rate, defined in terms of $t$,
  cannot be considered as the time translation of the birth rate.

If we assume that the systems with an observed luminosity $L$
  have a death rate which is specified by a probability mortality function  
$k(L)$,
  and that the birth of XRBs is a Poisson process subject to a birth rate at  
time $t$,
  we can construct a simple statistical/stochastic birth-death formulation
  to describe the time-dependent evolution of the XRB population.
The corresponding equation is
\begin{equation}
   {d \over {dt}}\ n(L,t) = f(L,t) - k(L)n(L,t)
\end{equation}
  (\cite{kwu-E1:wu01}, \cite{kwu-E1:wu02}).
As an approximation,
  we define $k(L)$ to be the inverse of the characteristic lifespan $t_{\rm x}(L)$, 
  such that $t_{\rm x}(L)\propto 1/L$ implies $k(L) = \beta L$,
  where $\beta$ is a constant.
For simplicity we neglect the delay in the formation of XRBs 
  with respect to the starburst at  time $t_{\rm a}$
  and assume that the birth function due to the starburst takes the form
  $\lambda(L)\delta(t-t_{\rm a})$.
Thus, we have $f(L,t) = \lambda_0(L) + a\lambda(L)\delta(t-t_{\rm a})$,
  where $\lambda_0(L)$ is the continuous component of the birth rate of XRBs 
  and $a$ is a parameter specifying the relative strength 
  of the starburst component.
As $f(L,t)$ and $t_{\rm x}(L)$ are now specified,
  equations (4) and (1) can be solved, 
  yielding $n(L,t)$ and $N(>L,t)$.

An important property of the solutions is that, at any time $t$,
  the starburst component of $n(L,t)$ has a cutoff
  at a critical luminosity $L_{\rm c} = [\beta (t-t_{\rm a})]^{-1}$.
Here the cutoff is exponential, whereas in the idealised deterministic model 
  considered above it is step-like.
The location of the cutoff of the two cases is, however, identical.
The cutoff causes an exponential luminosity break in the $N(>L)$ distribution. 
Including time-delay effects in the formulation 
  will not eliminate the luminosity cutoff
  but, instead, will cause $N(>L)$ to appear as a broken power law
  for a power-law form $\lambda(L)$ (Soria, private communication).
For the continuous component of $n(L,t)$
  there is no cutoff at $L_{\rm c} = [\beta (t-t_{\rm a})]^{-1}$.
The corresponding cutoff is at ${\tilde L}_{\rm c} \sim [\beta t_{\rm gal}]^{-1}$, 
  where $t_{\rm gal}$ is the age of the host galaxy.
The luminosity break is therefore robust,
  and it is a characteristic of aging of a population of XRBs born
  as a consequence of an episode of star-formation in the recent past.

We must emphasise that the results obtained from the above formulation
  should be interpreted in a probabilistic and statistical context.
The formulation allows the presence of systems with luminosity $L$
  brighter than the characteristic cutoff luminosity
  because the birth of XRBs is a Poisson process in the time domain
  and because the observed $L$ is an instantaneous quantity
  while the duration of the X-ray phase of an XRBs is related
  only to the luminosity averaged over the lifespan of the source.
Moreover, $n(L,t)$ is the expected value over an ensemble of sources
  with a distribution of ages which is not directly observable.
(For more details, see \cite{kwu-E1:wu02}.)

Nevertheless, the results obtained by the deterministic and stochastic formulations
  are qualitatively similar,
  as can be demonstrated using the following example.
Consider a simple birth-rate function,
   $f(L,t) =  \lambda(L)[ 1 + a \delta(t-t_{\rm a})]$,
   where $a$ is now the ratio of the strength of the impulsive starburst component 
   to that of the continuous formation component.
(Here we assume the same initial XRB brightness distributions
   for the two components.)
Then we have 
\begin{eqnarray}
  n(L,t) & = & n_o(L)~e^{-\beta L t} + {{\lambda(L)} \over {\beta L}}
        \biggl( 1-e^{-\beta L t} \biggr) \nonumber \\
      & & \   + {a \lambda(L)}
        \theta (t - t_{\rm a}) e^{-\beta L(t-t_{\rm a})} \ ,
\end{eqnarray}
   where $n_{\rm o}(L) \equiv n(L,0)$ is the initial population.
By comparing the last term in equation (5) with equation (3)
  (with $t_{\rm x}(L) = (\beta L)^{-1}$),
  we can easily see their similar time-dependent behaviours.
Also, for $\beta L t \gg 1$, the second term in equation (5)
  becomes $\lambda(L)/\beta L$, which is independent of $t$
  just as is the term ${\bar n}_0(L)$ in equation (3).  

Consider an initial population and a birth rate in power-law forms 
  with the lower cutoff below the limit of detection $L_*$, i.e.,
  $n_{\rm o} = n_{\rm o*} (L/L_*)^{-\alpha_1}$ 
  and $\lambda(L) = f_{\rm o*} (L/L_*)^{-\alpha_2}$, 
  where $\alpha_1$ and $\alpha_2$ are constants.
Then the cumulative luminosity distribution is
\begin{eqnarray}
  N(>L,t) & = &  n_{\rm o*} L_* (\beta L_* t)^{\alpha_1-1}
    \Gamma \big(1-\alpha_1, \beta L t \big)  \nonumber \\
    & &\  +\   {{f_{\rm o*}} \over \beta} (\beta L_* t)^{\alpha_2}~
      \biggl[{1 \over {\alpha_2}} (\beta L t)^{-\alpha_2}
          \biggl( 1 - e^{-\beta L t} \biggr)  \nonumber \\
     & &  \hspace*{0.85cm}   + \ {1 \over {\alpha_2}}
      \Gamma\big(1 -\alpha_2, \beta Lt \big) \biggr] \nonumber \\
    & & \ +\ {a f_{\rm o*}} L_*[\beta L_* (t-t_{\rm a})]^{\alpha_2-1}  
\nonumber \\
      & &\hspace*{0.85cm}
      \times \ \Gamma\big(1-\alpha_2, \beta L (t-t_{\rm a}) \big)
     \theta (t-t_{\rm a}) \ ,
\end{eqnarray}
   where $\Gamma (\alpha , x)$ is the incomplete gamma function.  

\begin{figure}[ht]
  \begin{center}
   \epsfig{file=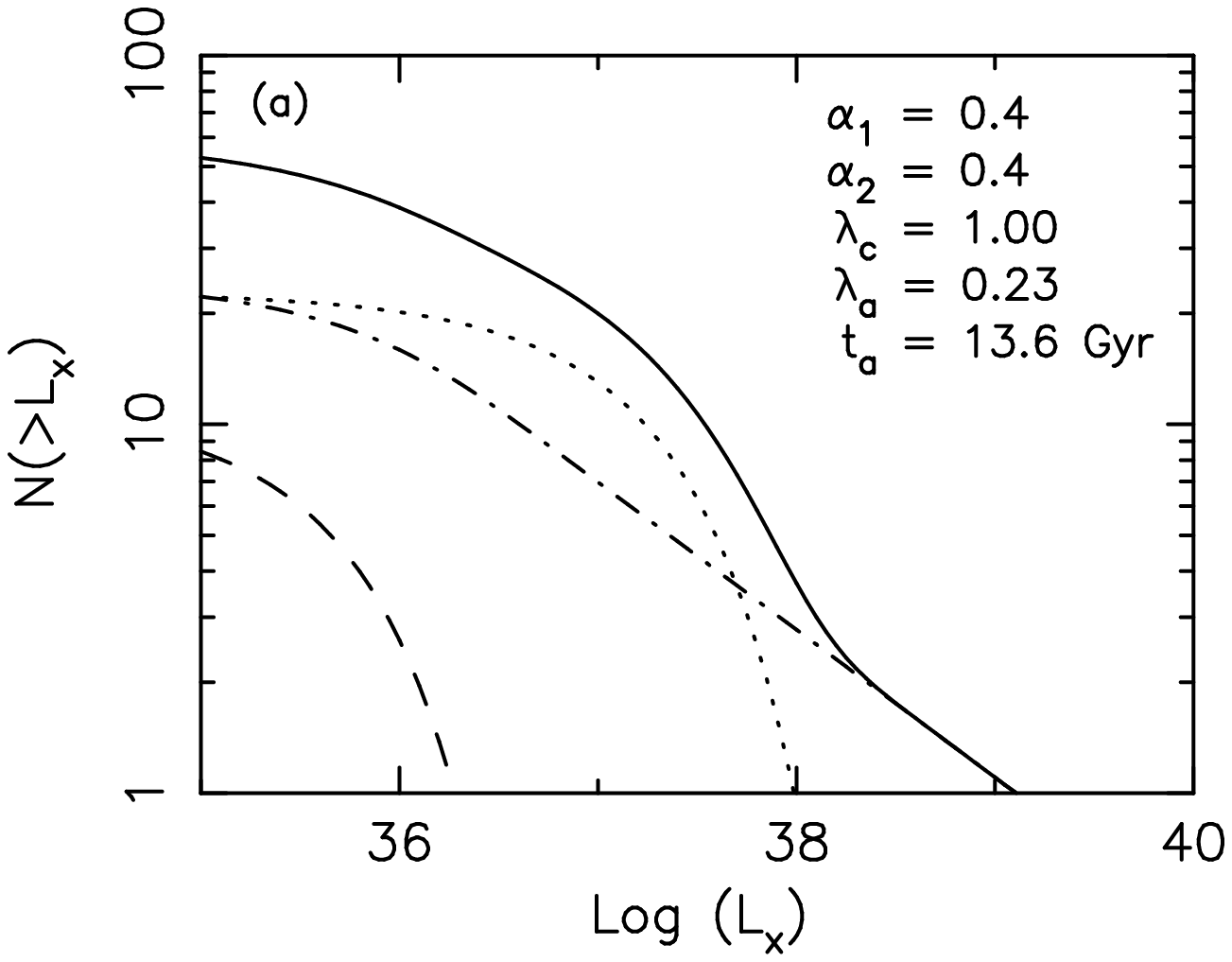, width=4.35cm}
   \epsfig{file=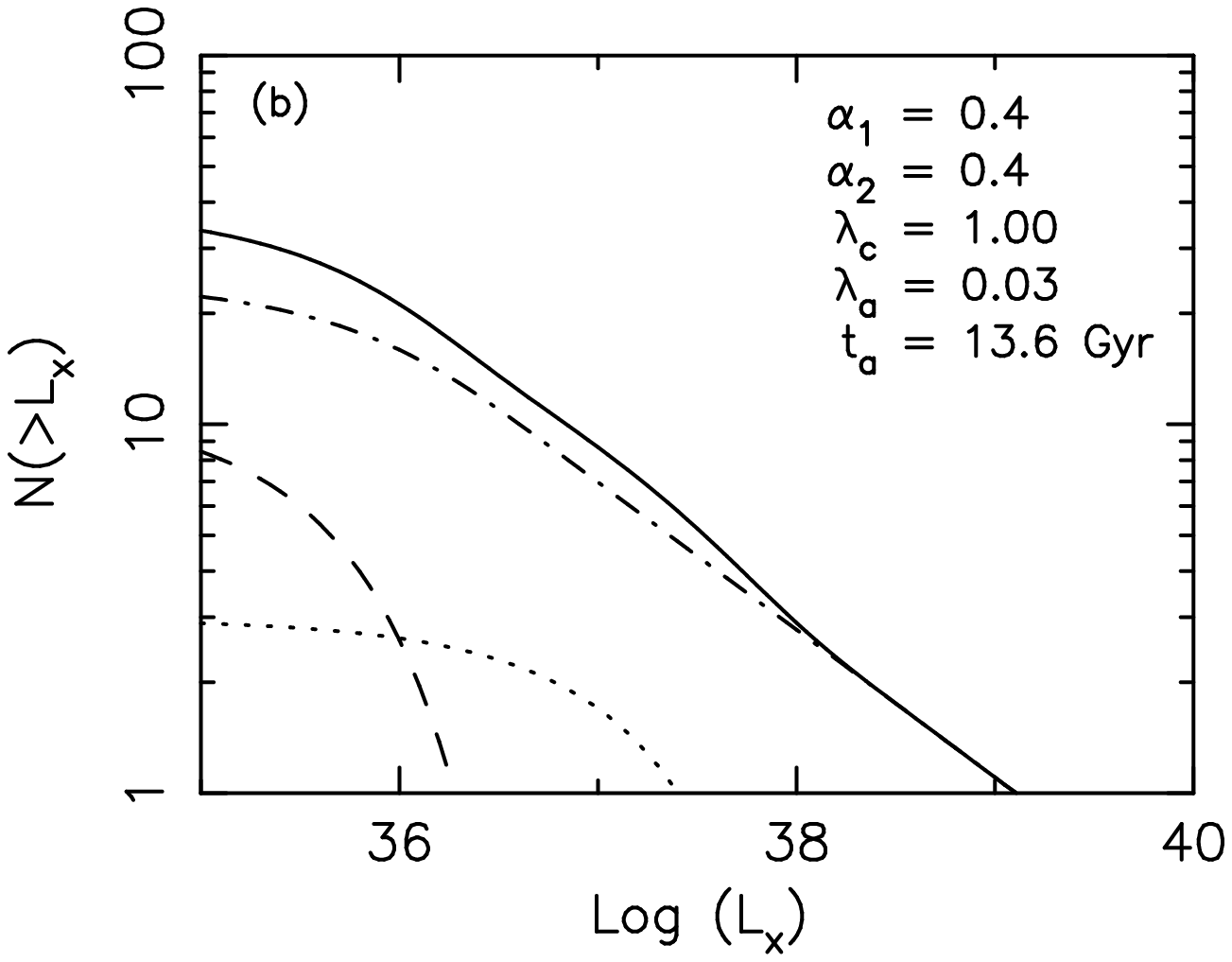, width=4.35cm}
   \epsfig{file=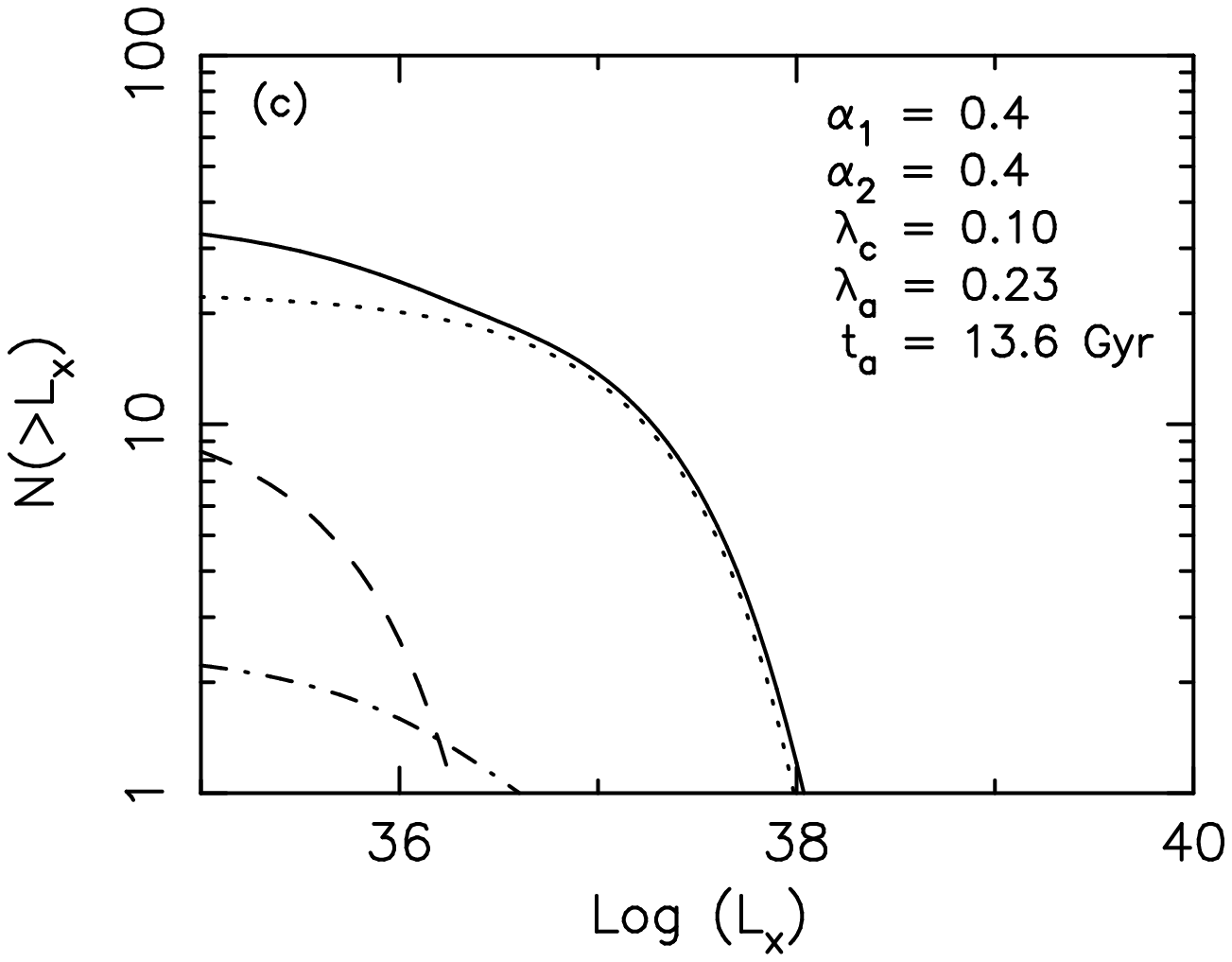, width=4.35cm}
   \epsfig{file=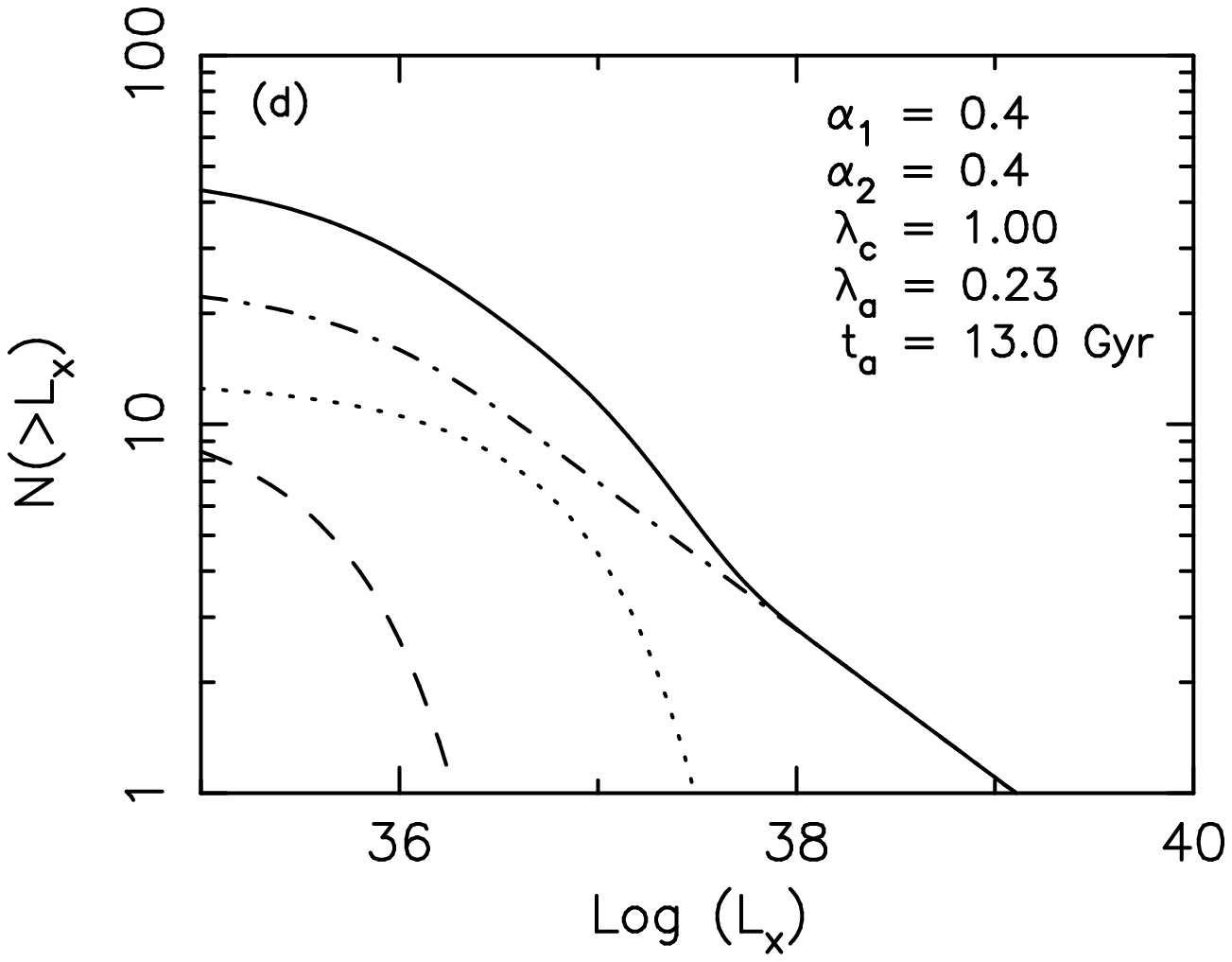, width=4.35cm}
  \end{center}
\caption{The $N(>L)$ curves of XRBs in a galaxy calculated 
    from the stochastic birth-death model
        without differentiating black-hole and neutron-star binaries.
     The parameters $\lambda_c \equiv (f_{o*}/n_{o*}\beta L_*)$
        and $\lambda_a \equiv (a f_{o*}/n_{o*})$
        specify the contributions of the continuous
        and the impulsive star-burst component respectively.
     The parameter $\beta$ is set to be $[0.3~{\rm M}_\odot c^2]^{-1}$,
        and the power-law indices $\alpha_1$ and $\alpha_2$
     are both fixed at 0.4.
   (a): The starburst episode occurred 400 million years ago
        ($t-t_{\rm a} =0.4$~Gyr).
     The values for $\lambda_c$ and $\lambda_a$ are
       1.0 and 0.23 respectively.
     The normalisation is chosen such that the source counts
       are similar to the number of the M81 bulge sources observed by Chandra 
       (see Tennant et al.\ 2001).
     The primordial component is represented by the dashed line;
       the continuous component by the dot-dashed line; and
       the impulsive starburst component by the dotted line.
     The total population is represented by the solid line.
   (b): Same as (a) except $\lambda_a = 0.03$.
   (c): Same as (a) except $\lambda_c = 0.1$.
   (d): Same as (a) except $t-t_{\rm a} =1$~Gyr.  }
\label{fauthor-E1_fig:fig2}
\end{figure}

Figure 2 shows four examples of $N(>L)$ calculated from this model.
The luminosity break is clearly present 
  for the impulsive-starburst component. 
By adjusting the relative strengths of the continuous and starburst components,
  we can obtain a form of $N(>L)$ 
  which is a single power law or which has a luminosity break.
Note the curve in panel (a) are very similar to
  what is observed in M81 (Fig.~3). 

\begin{figure}[h]
  \begin{center}
   \epsfig{file=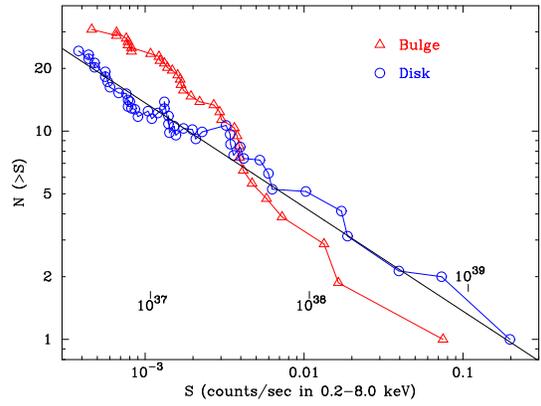, width=6.5cm}
  \end{center}
\caption{The (background substracted) log($N(>S)$)--log($S$) curves
   of the disk and bulge sources in M81 observed by Chandra.
   Adopted from Tennant et al.\ (2001).  }
\label{fauthor-E1_fig:fig3}
\end{figure}

\section{Discussion}

Above the completeness limit,
  the log($N(>S)$)--log($S$) curves of XRBs in individual nearby galaxies
  can be described as a single or a broken power law.
Separating the sources in the bulge and disk of the spiral galaxy M81
  reveals that the log($N(>S)$)--log($S$) curves of the XRBs
  in the two galactic components are different (\cite{kwu-E1:ten01}).
While the log($N(>S)$)--log($S$) curve of the disk sources is a single power law,  
  that of the bulge sources is a broken power law.
A power-law form for log($N(>S)$)--log($S$) curve is generally found for XRBs  
  in galaxies or galactic components with strong ongoing star-formation activity 
  (e.g., the disk of M101, \cite{kwu-E1:pen01};
  the nuclear region of M83, \cite{kwu-E1:sor02};
  and the starburst galaxy M82).
Broken power-law log($N(>S)$)--log($S$) curves are, on the other hand, found 
  for the sources in early-type galaxies (e.g.\ NGC~4697, \cite{kwu-E1:sar00}) 
  and the bulges of spiral galaxies (e.g.\ M31, \cite{kwu-E1:sup01})

\cite*{kwu-E1:sar00} proposed that
  the luminosity break is caused by a population of XRBs
  containing a neutron star which accretes at the Eddington limit.
Using the stochastic formulation described above
  and assuming the Eddington luminosity limit
  ($L \sim 2 \times 10^{38}~{\rm erg~s}^{-1}$) for the neutron-star XRBs,
  it can be shown that a prominent luminosity break can be created 
  (see \cite{kwu-E1:wu02}).
The break marks the Eddingtion limit for 1.5-M$_{\odot}$ compact objects.
As this break is stationary and has the same value for all galaxies,
  it can be used as a distance indicator (\cite{kwu-E1:sar00}).

While it is possible that the luminosity breaks in the log($N(>S)$)--log($S$) 
  curves of elliptical galaxies
   are caused by the luminosity limit of neutron-star XRBs,
  the luminosity breaks in the log($N(>S)$)--log($S$) curves
  of M31 (\cite{kwu-E1:shi01}) and of the bulge sources in M81  
  (\cite{kwu-E1:ten01})
  at about  $4 \times 10^{37}~{\rm erg~s}^{-1}$
  must be due to another mechanism.

From the birth-death model, we can see that a natural explanation 
  for the luminosity break is the aging of a populations of XRBs
  born at approximately the same time in the recent past.
If this explanation is correct, 
  then the location of the break marks the starburst epoch 
  and the luminosity function becomes an indicator of the age of the host galaxy 
  (or the host galactic component).  
The location of the break for the Chandra sources in the M81 bulge, for example, 
  indicates that $(t-t_{\rm a}) \approx 0.4~{\rm Gyr}$,
  which is consistent with the time M81 underwent a close encounter with its 
  neighbor M82 (\cite{kwu-E1:deg01})

\begin{acknowledgements}
We thank R. Soria for discussion. 
K.W.\ thanks M. Weisskopf for providing the funds for his visits to MSFC.
This work is partly supported by a University of Sydney Sesqui R\&D grant
  and a Royal Society travel fund.
\end{acknowledgements}

\end{document}